# AGENTS OF CHOICE: TOOLS THAT FACILITATE NOTICE AND CHOICE ABOUT WEB SITE DATA PRACTICES

Dr. Lorrie Faith Cranor
*Senior Technical Staff Member*
*AT&T Labs-Research*
*Shannon Laboratory*
*Florham Park, New Jersey, USA*
http://www.research.att.com/~lorrie/

## ABSTRACT

*A variety of tools have been introduced recently that are designed to help people protect their privacy on the Internet. These tools perform many different functions including encrypting and/or anonymizing communications, preventing the use of persistent identifiers such as cookies, automatically fetching and analyzing web site privacy policies, and displaying privacy-related information to users. This paper discusses the set of privacy tools that aim specifically at facilitating notice and choice about Web site data practices. While these tools may also have components that perform other functions such as encryption, or they may be able to work in conjunction with other privacy tools, the primary purpose of these tools is to help make users aware of web site privacy practices and to make it easier for users to make informed choices about when to provide data to web sites. Examples of such tools include the Platform for Privacy Preferences (P3P) and various infomediary services.*

## INTRODUCTION

As growing numbers of people use the Internet for an ever-expanding range of activities, the amount of personal data collected via the Internet is increasing. This phenomenon has raised a variety of privacy-related concerns, including:

- concerns about the secure storage and transfer of information;

- concerns that individuals' information may be collected without their knowledge or consent;

- concerns that the ease with which information can be collected and processed is leading to an increasing amount of data collection, database matching, and secondary use of data; and

- concerns that an individual's information may be transferred across jurisdictional boundaries to locations where it is not protected by the same privacy laws in effect where that individual resides.

Most of these concerns are not new; indeed they all existed well before the advent of the Internet. However, as the Internet becomes more pervasive these concerns are exacerbated.

A variety of legislative and regulatory efforts, self-regulatory programs, and new technologies have been launched in an effort to address online privacy concerns. Many of the regulatory and legislative efforts are not focussed on the Internet *per se*, but are part of more general efforts to institute new privacy laws. However, some recent legislation has focussed specifically on the Internet. For example, last year the United States Congress enacted the Children's Online Privacy Protection Act of 1998, which limits the ability of web sites to collect personal information from children under the age of 13. A number of self-regulatory efforts in the US are focussed specifically on the Internet, including the

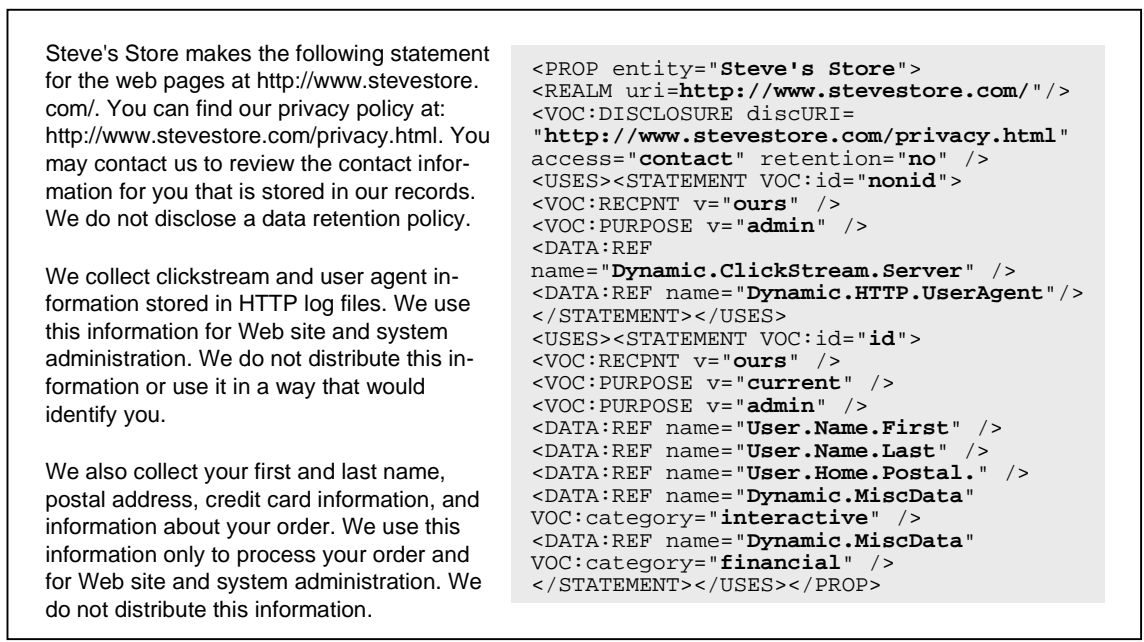

**Figure 1. A P3P privacy disclosure in English (left) and P3P syntax (right). This disclosure is based on a June 1999 draft of the P3P specification; the syntax may change in future drafts.**

Online Privacy Alliance and several online privacy seal programs.

In the technological domain, tools to protect online privacy perform many different functions. A variety of tools are available for encrypting files and email, establishing secure channels to web sites, and establishing encrypted "tunnels" between two computers on the Internet. These tools prevent eavesdropping and protect data from unauthorized access. In addition, anonymity tools are available that prevent online communications from being linked back to a specific individual and prevent eavesdroppers from learning with whom an individual is communicating [6, 11, 14]. Some tools allow users to build anonymous, yet persistent, relationships with web sites, thus allowing sites to track user behavior or provide customized services without building identifiable user profiles [5].

As popular web browsers have added features that make it easier for web sites to track an individual's browsing behavior, tools have been developed to disable these features. For example, a variety of tools are available that can block or give users more control over the use of web "cookies" – bits of information that a web site can store on a user's computer which will be automatically transmitted back to that site every time the user returns. In addition, many of the anonymity and cookie blocking tools also block the automatic transmittal of the "referer" field, which tells web sites the address of the last site the user visited. While this field can help sites learn how people are finding out about them and can be useful for tracking down problems, it also makes it easier for sites to build profiles of visitors and can sometimes be particularly dangerous when the address of the previous site includes confidential account information, credit card numbers, or search strings.

This paper discusses another set of privacy tools, agents of choice – the set of privacy tools that aim specifically at facilitating notice and choice about Web site data practices. While these tools may also have components that perform other functions such as encryption, or they may be able to work in conjunction with other privacy tools, the primary purpose of these tools is to help make users aware of web site privacy practices and to make it easier for users to make informed choices about when to provide data to web sites.

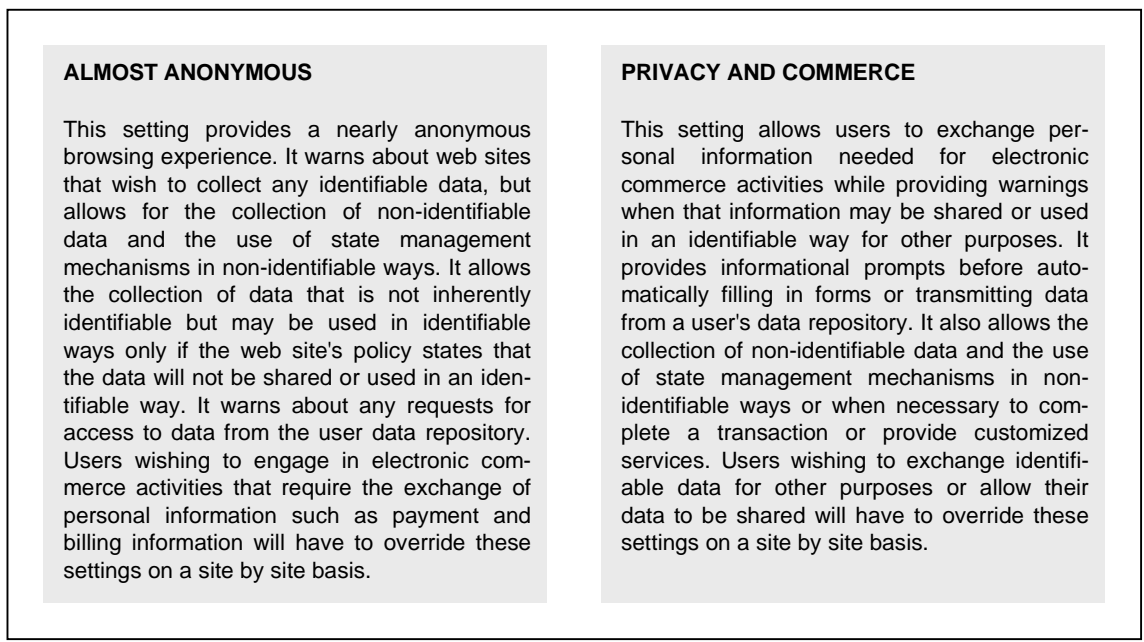

**Figure 2. Two examples of the kinds of settings APPEL files might provide.**

## THE PLATFORM FOR PRIVACY PREFERENCES (P3P)

The World Wide Web Consortium (W3C) launched the Platform for Privacy Preferences (P3P) project to develop a standard way for web sites to communicate about their data practices. The goal was to enable machine-readable privacy disclosures that could be retrieved automatically by web browsers and other user agent tools. These tools would then compare each disclosure against the user's privacy preferences and assist the user in deciding when to exchange data with web sites. Unlike anonymity tools, which seek to prevent any transfer of personally-identifying information, the P3P effort assumes that there are some situations where users desire to reveal personal information. Thus the P3P activity seeks to enable the development of tools for making informed decisions about when personal information should be revealed.

The P3P specification includes a standard "vocabulary" for describing a web site's data practices, a set of "base data elements" that web sites can refer to in their P3P privacy disclosures and explicitly request from the user, and a protocol for requesting and transmitting web site privacy disclosures and data [9,10]. Figure 1 shows an example web site privacy disclosure in both the machine-readable P3P syntax and in English.

P3P also includes a standard language for encoding a user's privacy preferences called A P3P Preference Exchange Language (APPEL). APPEL files specify what actions the user agent should take depending on the type of disclosures made by a web site. The four standard actions defined in the APPEL specification are seamlessly agreeing to a site's practices, providing an informational prompt to the user, warning

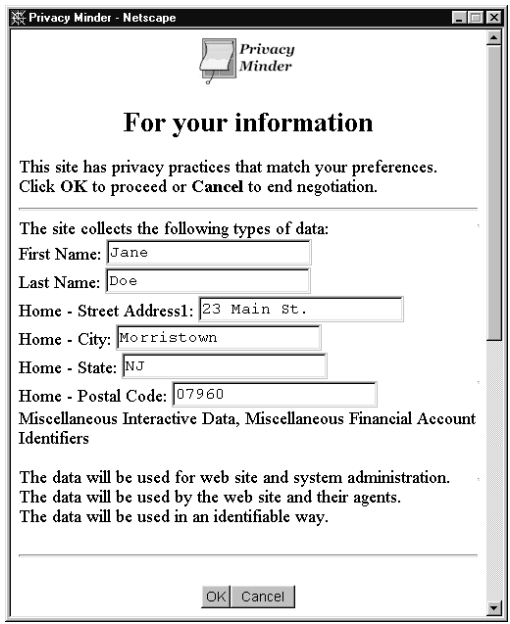

**Figure 3. Example of form automatically generated by the Privacy Minder 1.0 P3P user agent software.**

the user, and seamlessly rejecting a site's practices.

P3P user agent implementations are expected to include interfaces for users to specify their preferences about data usage. These interfaces may allow users to import "canned" APPEL files that match their preferences. Such files might be distributed by privacy advocacy groups, privacy seal providers, governmental privacy agencies, or other organizations that users trust. Figure 2 shows two examples of the kinds of settings these APPEL files might provide.

A P3P user agent might also include a "user data repository" where users can store data that they frequently exchange with web sites. The data in this repository is identified by the standard names defined in the P3P base data set. (P3P also includes a mechanism for defining new data sets that can be referred to in privacy disclosures and stored in user data repositories.) Users who never or rarely wish to provide data to web sites might choose not to enter data into their repositories; however, users who frequently exchange data with web sites may find it convenient to store data in their repositories. In addition, use of the standard base data set elements names creates a tight coupling between the privacy disclosure and the data being transferred, reducing the ambiguity about the kinds of data to which a site's practice disclosures apply. For example user agent implementations might automatically create and fill out forms with repository data elements when sites request them, annotating the forms with the site's data practices. Figure 3 shows an example of such a form generated by the Privacy Minder 1.0 user agent software. Privacy Minder is a prototype P3P user agent developed at AT&T Labs-Research (see http://www.research.att.com/projects/p3p/pm/).

In addition to automating the decision-making process, P3P user agents can also provide tools that make it easier for users to quickly assess a site's privacy practices for themselves. The informational prompt shown in Figure 3 is one such mechanism. In addition user agents might display symbols that summarize a site's privacy policy or indicate that it has a privacy seal or is bound by certain privacy laws. They might also include buttons that users can click to jump directly to a site's privacy policy disclosure without having to search for it on the site. Figure 4 shows some of the tools provided by the Privacy Minder 1.0 software.

**INFOMEDIARIES**

Since the beginning of 1999, at least five companies have announced new services and tools that help people manage their on-line identities and protect their privacy. These services and tools are often referred to as *infomediaries*, using the term coined by John Hagel [7].

Five infomediaries are described below. Most had only released demonstration or prototype products as of June 1999. Almost all say they are either based on P3P or have plans to use P3P after the specification is finalized. However, few details were available about how they would use P3P. Most of these infomediaries allow users to store information in secure personal data stores and use it in conjunction with automatic form filling features. Some restrict automatic form filling to sites that have policies that match a user's privacy preferences. Some also have mechanisms that allow users to opt-in to automatically sharing information with marketers of products or services they have expressed interest in -- sometimes anonymously; sometimes in exchange for discounts, coupons, or monetary compensation.

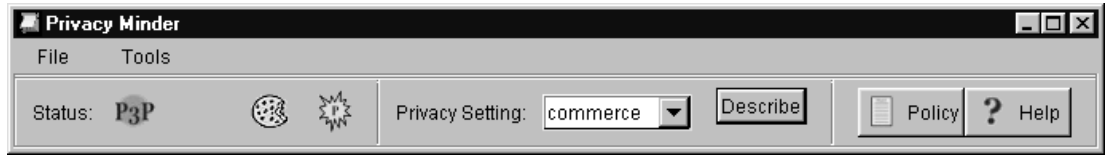

**Figure 4. The Privacy Minder 1.0 toolbar includes icons to indicate a P3P-enabled site, the use of cookies at the site, and the presence of a privacy seal [1]; the "Policy" button allows users to jump directly to a site's privacy policy.**

**digitalme**
http://www.digitalme.com

The Novell *digitalme* technology will allow users to create various identity "cards" that can be shared on the Internet according to users' preferences. Users can control what information is stored in each card and the conditions under which it may be shared.

**Jotter**
http://www.jotter.com

Jotter Technologies offers a free "personalized desktop toolbar" called *Jotter*. In addition to automated form filling functions and other features, the Jotter tool bar includes a privacy button that can be dragged onto a browser window, causing Jotter to attempt to automatically locate the privacy policy of the page currently shown in that window.

**Lumeria**
http://www.lumeria.com

Lumeria has developed an infomediary platform that uses the company's *SuperProfile* personal profiling technology. The SuperProfile allows users to store their personal information securely on the Internet and release it according to the user's preferences. The SuperProfile provides options for users to control the circumstances under which anonymous and/or identified profile information is released. Users of the SuperProfile may be able to earn money when they choose to release their personal information.

**PrivacyBank.com**
http://www.privacybank.com

PrivacyBand.com offers a free *AutoFill* service. AutoFill allows users to store their personal information and privacy preferences in the PrivacyBank database. When a user visits a supported web page that matches their privacy preferences, she can click an Auto Fill button to have the form automatically filled out.

**Privaseek**
http://www.privaseek.com

PrivaSeek Inc. is currently beta testing their *Persona, Valet,* and *Vault* tools. These tools allow users to store their personal information securely and automatically fill out forms on web sites that request information stored in the user's Privaseek Persona. The Persona may be customized to automatically share or sell information to pre-screened web sites that market products or services a user is interested in. These sites must agree to treat information collected from users according to their privacy preferences.

**DISCUSSION**

The introduction of privacy tools that facilitate notice and choice about Web site data practices has been greeted with both praise and criticism. Critics suggest that while these tools may increase users' knowledge about web site privacy practices, they do little or nothing to ensure that web sites actually have policies that protect users' privacy [2]. Proponents of choice tools argue that such tools can help enforce privacy directives. For example, P3P users might install APPEL files that reject privacy policies that do not comply with their country's privacy laws. In addition, choice tools may eventually result in the creation of markets that allow users to choose how much information to provide to web sites. A recent Harvard Law Review analysis of P3P concluded [8]:

> …the multitude of potential substitutes for any particular type of Internet content, coupled with the intense competition among content providers for Internet traffic, ensures a high level of site responsiveness to user preferences…. A P3P regime will result in the optimal level of privacy protection because it permits individuals to value privacy according to their personal preferences. Individual users will configure their privacy preferences to protect privacy according to the value that they attach to it. In the resulting privacy market, those who value their personal information less will part with it more easily than those who value it more…. When aggregated, these individual preferences will exert pressure on site operators to conform their privacy practices to user preferences.

While many of the developers of choice tools remain hopeful that once users have the ability to easily obtain and respond to web site privacy policies sites will be pressured to adjust their policies to meet user preferences, this analysis may be overly optimistic about the market mechanism. It remains to be seen whether businesses will offer privacy choices that meet users' preferences or whether individuals will lower their expectations about privacy to match the choices offered in the marketplace [10].

Some privacy advocates warn against the commoditization of privacy [4] and argue that choice tools will ultimately lead to more disclosure of personal data [12]. They suggest that mechanisms such as automatic form filling make it easier for users to provide data and compensation mechanisms will tempt users to sell their data [2].

Most of the choice tools have been designed with the assumption that both users and web sites may benefit from the sharing of data. The infomediaries, in particular, seem to be oriented towards encouraging users to share data. Some appear to provide privacy-enhancing value only to users who activate automatic data sharing features, and in some cases only at sites that use the infomediary's proprietary technology. On the other hand, many of the infomediaries are designed to enable marketers to collect data through opt-in mechanisms and build pseudonymous relationships where possible instead of using more privacy-invasive traditional marketing techniques.

P3P has been designed to provide value to users regardless of whether they take advantage of automatic data sharing features. Indeed these features are optional parts of the P3P specification [9] and the P3P Guiding Principles [3] states that P3P enabled sites should "limit their requests to information necessary for fulfilling the level of service desired by the user." In fact, P3P has been criticized by some marketers for being likely to "precipitate a decrease in the flow of marketing information, even where the intended use is benign." Two Citibank employees raised this and other concerns about P3P in a white paper (not necessarily representative of official Citibank policy) last year [13]:

> P3P allows a user to dictate under what sort of conditions she is willing to give out personal information. If Citibank does not agree to whatever conditions the user puts forth, the user may opt to not transact with the bank at all – thus putting the onus on the bank to tighten the privacy protection until users are willing to transact i.e., to the lowest common denominator.
>
> There is a concern that P3P would let ordinary users see, in full gory detail, how their personal information might be misused by less trusted or responsible web site operators. Such knowledge may cause users to resist giving out information altogether. Some individual business groups have done focus studies on users, and … concluded that most users would prefer to give out only information needed for the transaction and that they do not like the idea of someone monitoring their browsing behavior.

Clearly the jury is still out on what impact choice technologies will actually have on web site data practices and user behavior. The impact of these tools will depend on a variety of factors including how they are ultimately implemented, how Internet users value their privacy, and how well these tools complement legal and regulatory frameworks.